\documentclass[preprint,11pt]{aastex}
\usepackage{natbib}
\usepackage{epsfig}
\usepackage{multirow}
\makeatletter
\renewcommand\@biblabel[1]{\hspace{-\labelsep}}
\makeatother

\begin{document}

\title{Constraining the Orbit of Supermassive Black Hole Binary 0402+379}

\author{K. Bansal\altaffilmark{1}, {G. B. Taylor\altaffilmark{1}},
{A. B. Peck\altaffilmark{2}}, {R. T. Zavala\altaffilmark{3}}, \& {R. W. Romani\altaffilmark{4}}
}

\altaffiltext{1}{Department of Physics and Astronomy, University of New Mexico, Albuquerque, NM 87131}
\altaffiltext{2}{National Radio Astronomy Observatory, 520 Edgemont Rd, Charlottesville, VA 22903}
\altaffiltext{3}{United States Naval Observatory, Flagstaff Station 10391 W. Naval Observatory Rd. Flagstaff, AZ 86001}
\altaffiltext{4}{Department of Physics, Stanford University, Stanford, CA 94305-4060}

\begin{abstract}
The radio galaxy 0402+379 is believed to host a supermassive black hole binary (SMBHB). The two compact core sources 
are separated by a projected distance of 7.3 pc, making it the most (spatially) compact resolved SMBHB known. We present 
new multi-frequency VLBI observations of 0402+379 at 5, 8, 15 and 22 GHz, and combine with previous observations 
spanning 12 years. A strong frequency dependent core shift is evident, which we use to infer magnetic fields
near the jet base. After correcting for these shifts we detect significant relative motion of the two cores
at $\beta=v/c=0.0054 \pm 0.0003$ at $PA= -34.4^\circ$. With some assumptions about the orbit,
we use this measurement to constrain the orbital period $P\approx 3 \times 10^4$ y and 
SMBHB mass $M \approx 15 \times 10^9\ M_\odot$.  While additional observations are needed to 
confirm this motion and obtain a precise orbit, this is apparently the first black hole system resolved as a visual
binary.
\end{abstract}

\section{Introduction}


It is commonly believed that the later stages of galaxy evolution are
governed by mergers. It is very common for galaxies to collide and
interact with each other. Considering that most galaxies in the
universe harbor supermassive black holes (SMBH) at their centers
\citep{rich98}, it can be inferred that massive black hole pairs
should be the outcome of such mergers through the hierarchical
formation of galaxies \citep{beg80}. This implies that SuperMassive
Black Hole Binaries (SMBHB) should be relatively common in the
universe. However, despite very extensive searches, very few such
systems have been observed \citep{burke, tremb}. The reason for this
could be that black holes in a binary system either merge quickly, or
that one of them escapes the system \citep{merritt05}. Hence, understanding these SMBHB
systems is important to understand a variety of processes ranging from
galaxy evolution to active galactic nuclei (AGN) to black hole growth.

There are two types of galaxy mergers, major and minor. Major mergers result when the interacting galaxies are of similar sizes (mass ratio less than 3:1 (\citet{stew09}, and references therein)), whereas in the case of minor mergers one galaxy is significantly larger than the other. A crucial expectation related to galaxy mergers is the emission of gravitational waves. When galaxies merge, due to the dynamical friction between them, the black holes at their corresponding centers sink towards a common center. This leads to the formation of a binary system, such that its orbit decays due to the interaction between the stars, gas, and dust of both galaxies. The two black holes may reach a small enough separation that energy losses from gravitational 
waves allow the binary to coalesce into a single black hole \citep{beg80,mm03}. 

Numerous simulations have been performed to study these SMBH mergers. These simulations deal with various aspects such as black hole mass ratio, self or non-gravitating circum-binary discs, orbital spin, black hole spins, gas or stellar dynamics etc. \citep{barnes, merritt05,escala04,escala05,sesana06, dotti07, cal09, cal11,khan11,schnit13}. In spite of several attempts, attaining the required resolution (last parsec problem $\sim 0.01$ pc) to study the black hole coalescence has been challenging. The fate of the merger at the parsec scale depends on the amount of surrounding stars and gas, and their interaction with the binary. In the case of a stellar background, due to 3-body scattering, formation of loss cone takes place \citep{sesana07}, whereas, in the case of gas rich mergers tidal forces inhibit the gas from falling onto the binary, hence creating a gap (\cite{dotti12}, and references within). The loss cone causes a decay period longer than the Hubble time \citep{mm03,merritt07}, and hardening can be attained for a triaxial stellar remnant with the loss cone being replenished \citep{merritt11} and an expected coalescence time of $\sim 10^{8}$ years \citep{khan11}. For gas-rich mergers, the gap doesn't inhibit gas flow \citep{roedig12} and a massive circumbinary disc around the binary promotes the decay leading to a timescale for an equal mass binary $\sim 10^{7} M_{\odot}$ that is less than the age of the Universe \citep{hayasaki09}. For higher mass black holes ($\sim 10^{8-9} M_{\odot}$) the timescales are greater. In a recent study by  \citet{khan16}, they report that for massive galaxies at high redshifts ($z>2$) it takes about few million years for black holes to coalesce once they form a binary, whereas, at lower redshifts where nuclear density of host is lower, it may take longer time of order a Gyr.


Gravitational waves from merging black holes are expected as a result of Einstein's General Theory of 
Relativity \citep{ae16,ae97,ae18,ae02}\footnote{http://einsteinpapers.press.princeton.edu/}. In September 2015, the Laser
Interferometer Gravitational Wave Observatory (LIGO) discovered a
gravitational wave (GW) source GW150914, and identified it as a
merging binary black hole (BBH) (Abbott et al. 2016a).
Although the masses of the two BHs are much smaller ($\thicksim$30 $M_{\odot}$)
in comparison to SMBHBs ($\thicksim$$10^{7} - 10^{10} M_{\odot}$),
this discovery provides the first observational evidence for the
existence of binary BH systems that inspiral and merge within the age
of the universe. It motivates further studies of binary-BH formation
astrophysics, and with the upcoming detectors such as evolving Laser
Interferometer Space Antenna (eLISA, \cite{elisa}), it will be
possible to detect low-frequency GW (around one mHz), emitted from the
inspiral of massive black holes \citep{elisa2}. While mergers of
SMBHB's are expected to be common emitters of GW radiation, modulating
pulsar timing observations have not yet detected any evidence for a GW
signal \citep{arzou16}. Pulsar timing observations, unlike LIGO, should be more
sensitive to SMBHB mergers \citep{shan15}. 0402+379, with a separation
of 7.3 pc between its core components, is one of the most important
precursors of GW sources, and is important to understand the reason
behind the low incidence of such systems. From the elliptical
morphology of the 0402+379 host galaxy \citep{andra16}, we believe
this object to be the result of a major merger.

\begin{figure}[ht]
\centering
\includegraphics[width=\textwidth,height = 0.9\textwidth,angle=0]{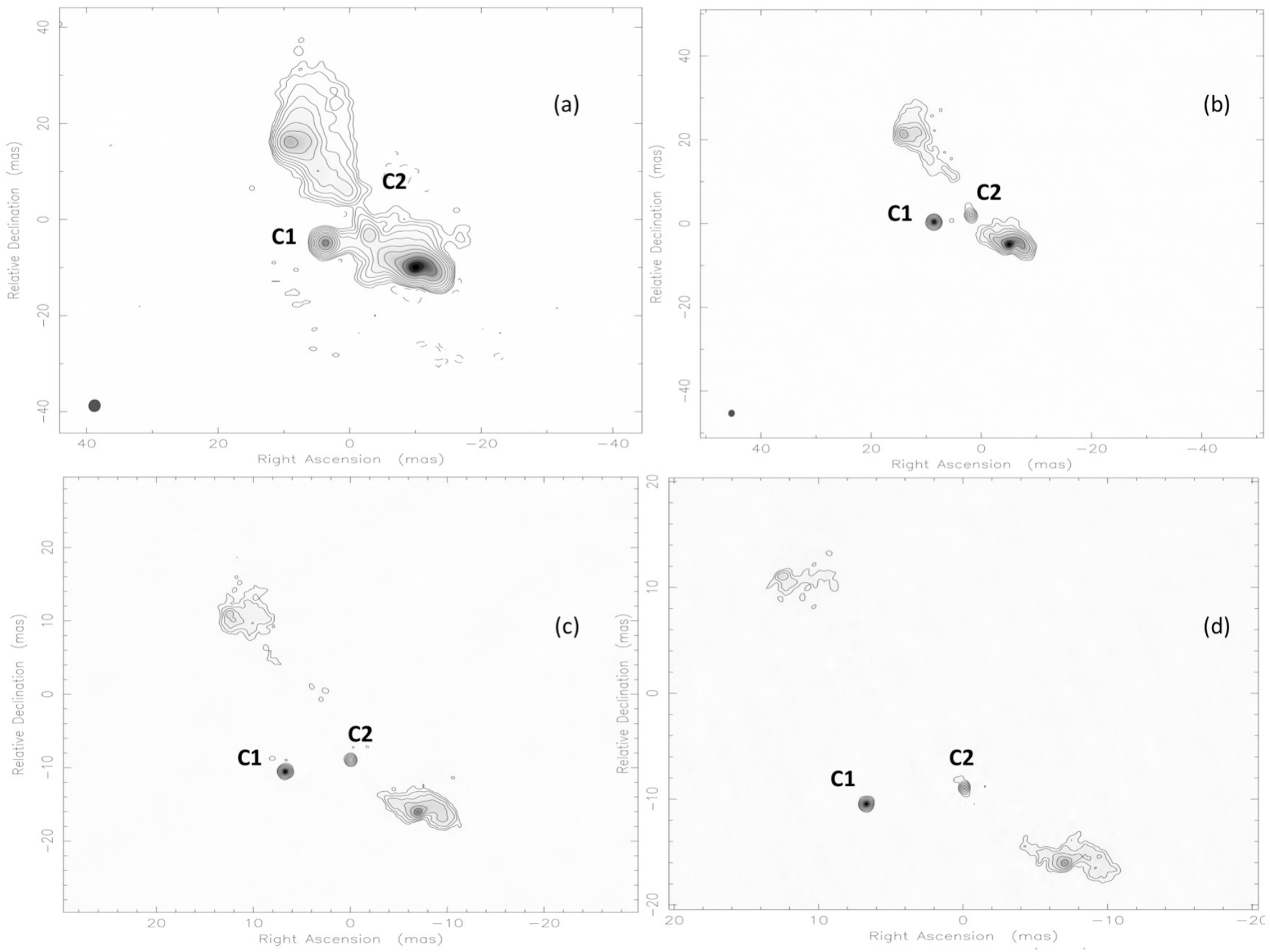}
\vspace{-0.5cm}
\caption{Naturally weighted 2015.43 VLBA maps of 0402+379 at 5, 8, 15 and 22 GHz. Designated C1 and C2, are the core components in 0402+379 \citep{man03,Rodri06}. Contours are drawn beginning at 0.15 $\ \sigma$ (a), 1 $ \sigma$ (b), 1 $\sigma$ (c) \& 1.5 $\sigma$ (d), and increase by a factor of 2 thereafter.
(a) Note that the core components are slightly resolved here. There is a bridge between these two components, and we believe this is a jet emanating from C1, as has been discussed in this paper. (b) A jet emerging from C2, moving in the direction of hotspots can be identified here clearly. We have used this map to obtain the jet-axis angle. (c) A very faint jet emanating from C2, similar to 8 GHz map, can be seen here. (d) No jets are visible at this frequency.}
\label{maps1}
\end{figure}

The radio galaxy 0402+379 was first observed by \cite{Xu95} as a part
of the first Caltech Jodrell Bank Survey (CJ1), although at that time
it was not identified as a SMBHB. This source first acquired attention
as a Compact Symmetric Object (CSO) candidate (small AGN with jets oriented
close to the plane of the sky such that the radio emission from the
jets is detected on both sides of the core) in 2003, in the full
polarimetry analysis by \cite{pol03}. Subsequently, \cite{man03}
studied this source at multiple frequencies using the VLBA\footnote{The National Radio
Astronomy Observatory is operated by Associated Universities, Inc.,
under cooperative agreement with the National Science
Foundation.} (Very Long
Baseline Array), and on the basis of its properties, they classified
it to be an unusual CSO. \cite{Rodri06} studied this source in more
detail and arrived at the conclusion that it is a SMBHB. This source
contains two central, compact, flat spectrum and variable components
(designated C1 and C2, see Fig.~\ref{maps1}), a feature which has not
been observed in any other compact source. This is one of the only
spatially resolved SMBHB candidates \citep{git13, dean14}. The
milliarcsecond scale separation requires high resolving power available
only with a telescope such as the VLBA. Although other systems like RBS 797 and J1502+1115 (a
triple system) with a separation of about 100 pc have been detected,
no system other than 0402+379 has been resolved at parsec scales. We
believe that this SMBHB is in the process of merging.

\cite{Rodri06} imaged this source at multiple frequencies, studying
the component motion at 5 GHz, but finding no significant detection of
core displacement. In this paper, we incorporate new 2009 and 2015 epochs of
0402+379 VLBA observations at 5, 8, 15 and 22 GHz, while re-analyzing
the 2003 and 2005 observations. These data show strong evidence for
a frequency dependent core-shift effect \citep{lob98, soko11}. After
accounting for this effect, our data set allows a detection of the
relative motion of the two cores, making this the first visual SMBHB.
We comment on the implications for the orbital period and masses.
Throughout this discussion, we assume $H_{0}=71$ kms$^{-1}$Mpc$^{-1}$
so that 1 mas = 1.06 pc.

\section{Observations and Data Reduction}

\subsection{VLBA Observations}

Observations were conducted on December 28, 2009 and June 20, 2015
with the VLBA at 4.98, 8.41, 15.35, and 22.22 GHz. For the 2015 
observations, the total time on
source was 70 min at 5 GHz, 260 min at 8 GHz, 290 min at 15 GHz, and
330 min at 22 GHz. 3C84 and 3C111 were observed for bandpass and gain
calibration, respectively. The data recording rate was 2048 Mbps with
two bit sampling. Each frequency was measured over eight intermediate
frequencies (IFs) such that every IF consisted of a bandwidth of 32 MHz
across 64 channels in both circular and their respective cross
polarizations.

Standard data reduction steps including flagging, instrumental time
delay, bandpass corrections, and frequency averaging were performed
with the NRAO Astronomical Image Processing System (AIPS)
\citep{van96, Ulvestad01}.  For all iterative self-calibration methods
the initial model was a point source. Further cleaning, phase and
amplitude self-calibration were executed manually using Difmap
\citep{Shep95}. The source structure was later model-fitted in the
visibility $(u, v)$ plane with Difmap using circular and elliptical
Gaussian components.

Fully calibrated VLBI archival data from the 2003 epoch \citep{man03}
at 15 GHz, and the 2005 \citep{Rodri06} epochs
at 5, 8, 15 and 22 GHz, have been included to study the core component
motion with frequency and time. The visibilities were imaged and model
fitted in Difmap, as with the 2009 and 2015 data, to obtain the core
positions. The calibrator 3C111 has been observed in the same configuration across all four epochs. Details of the observations can be found in Table~1.

\section{Measurement and Fits}

The new 2015 data set provides the most sensitive measurement of the
source geometry. Using model fitting in Difmap to the visibilities we determine the
Gaussian size, axis ratio and position angle for each core
component. These measurements are listed in Table 2. Additional Gaussian
components are
included in each model to account for the extended structure.  We then
follow \cite{Rodri06} and \cite{man03} in fixing these parameters in
fits to the other epochs, while allowing only the positions and fluxes
of C1 and C2 to vary. These are listed in Table 3, with the relative
positions listed as angular separation $r$ and position angle $\theta$
measured north through east.  The reported flux density errors combine the map 
rms $\sigma_{\rm
  rms}$ and an estimated systematic error in quadrature: $\sigma_S =
[(0.1 S_\nu)^2 +\sigma_{\rm rms}]^{1/2}$.

The effective position errors are more difficult to estimate. The
statistical errors $\approx a/2(S/N)$ \citep{fomalont} are very small ($\sim 2\ \mu$as),
and systematic effects certainly dominate.  We made an initial check
on these errors, by re-doing the previous Stokes I map analysis in
Stokes LL and RR. 0402+379 is an unpolarized source so we expect data from both Stokes RR and LL to yield similar distance measurements. Another way to obtain these error estimates is to split the data in time or frequency. However, these maps would have a different (u,v) coverage, thus making it difficult to make a comparison between them. If the polarization-dependent structure differences
are small as expected, then any measured differences can be attributed
to systematic errors. This technique has been discussed previously by \cite{roberts91} and \cite{mcgary01}. Decomposing the relative positions into RA($x$) and
DEC($y$), we find that the median shifts in the core centroid positions
are $\sigma_{x} = 7\ \mu$as and $\sigma_{y} = 8\ \mu$as for the higher
frequencies, and $\sigma_{x} = 31\ \mu$as and $\sigma_{y} = 34\ \mu$as at
5 GHz.


\subsection{Analysis}

\begin{figure}
\centering
\includegraphics[width = \textwidth, height = 0.8\textwidth]{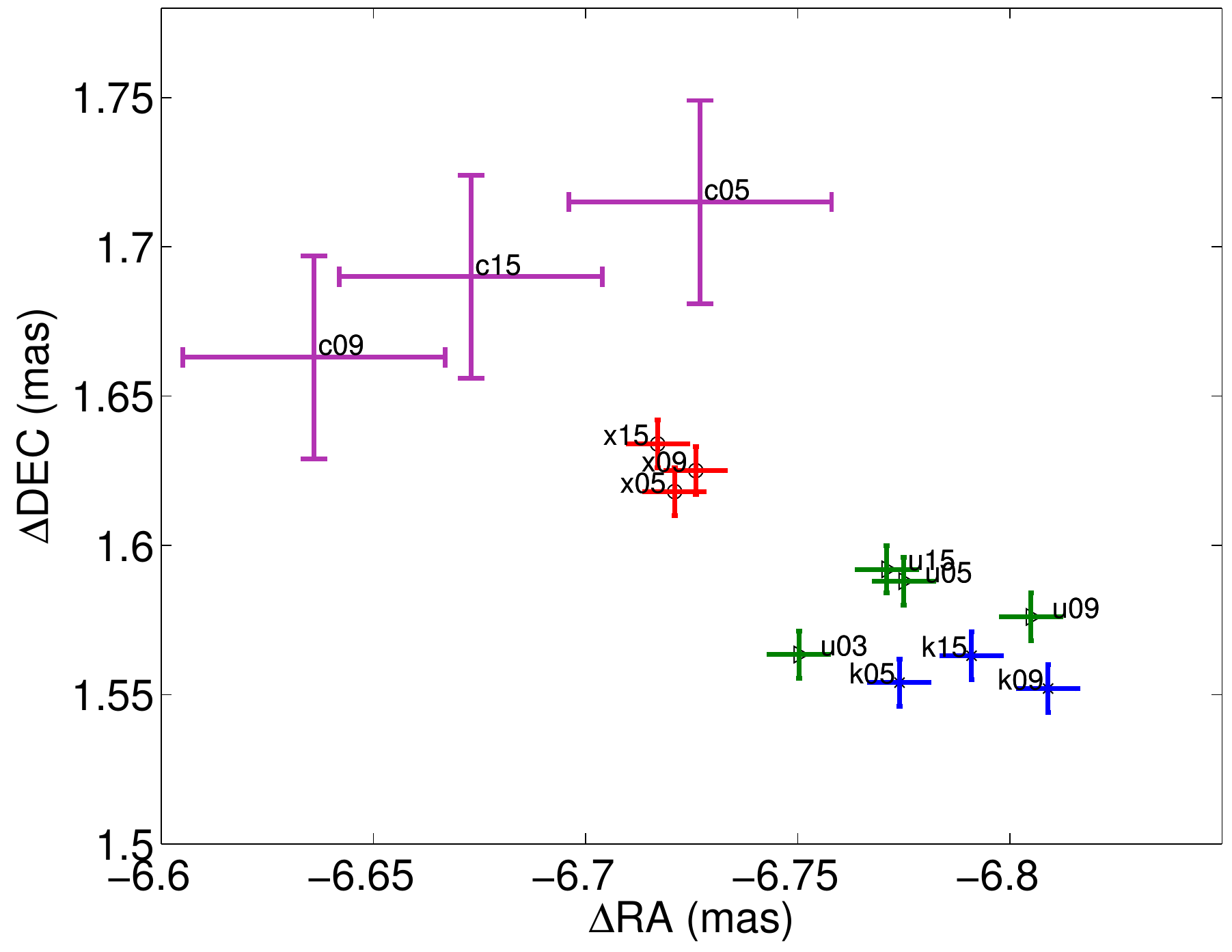}
\vspace{0.5cm}
\caption{We have plotted projected relative {\rm RA} {\it vs.} {\rm DEC} of component C2 with respect to C1 (at origin), at 5 GHz (c's), 8 GHz (x's), 15 GHz (u's), and 22 GHz (k's). This is the raw, uncorrected, modelfit positions. An 
offset in position with frequency can be seen due to the core-shift effect 
discussed in the text. }
\label{raw}
\end{figure}

The core component flux density arises from the surface at which the
self-absorption optical depth is unity \citep{bland79}. Since this is
strongly frequency dependent, we expect an asymmetric extended
structure, such as the jet base which defines the core, to have a
frequency-dependent centroid.  This effect is very obvious in the raw
positions (Fig.~\ref{raw}), where the lower frequency centroids are shifted
to the NE along the larger scale outflow position angle
(Fig.~\ref{maps1}a).  Similar shifts have been detected in a number of
AGN \citep{lob98, soko11}.  Since we are measuring position relative
to C1, any extension to that source may also contribute to the
relative core shift; this need not be at the same position
angle. However the combined shift appears to be dominated by C2 and we
indeed find that the frequency-dependent shift is along the $47^\circ$
(N-E) position angle of the C2 outflow. According to \citet{lob98}
the shift can be parameterized as $r_{c} = a \nu^{-1/k}$, where $a$ is the
shift amplitude and $k$ depends on the jet geometry, particle
distribution and magnetic field. For example a conical jet with a
synchrotron self-absorbed spectrum gives $k=1$ \citep{lob98}.  

\begin{figure}
     \begin{center}
            \includegraphics[width=\textwidth, height = \textwidth]{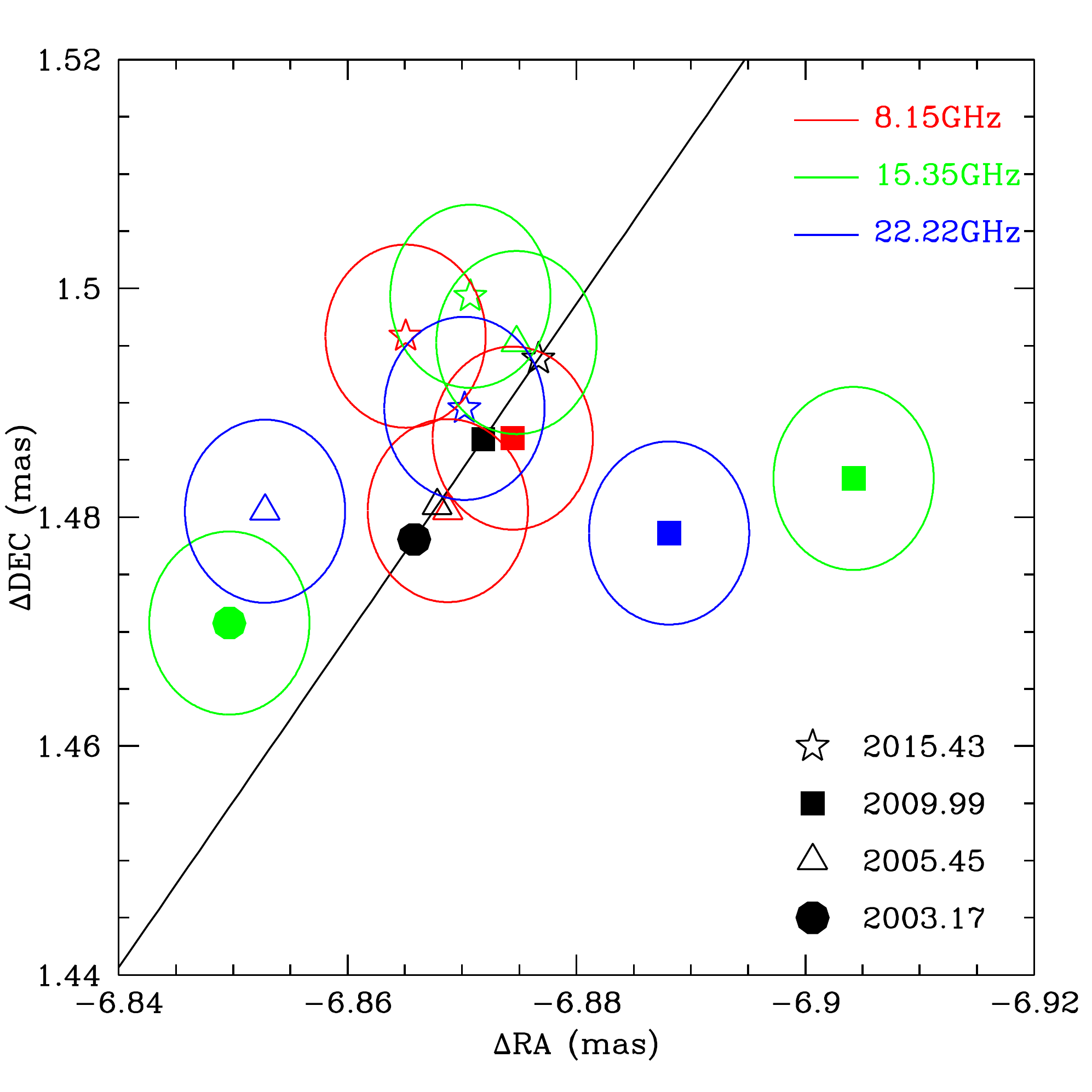}
    \end{center}
     \vspace*{-1.0cm}
    \caption{Position of C2 relative to C1 in time after removing the effect of the core shift.  The black line is a proper motion fit; the best fit positions at each epoch are labeled by points along the line.}
   \label{motions}
\end{figure}

By correcting to an infinite frequency we can mitigate the effect of this
core shift on the position of the nucleus (Fig.~\ref{motions}). We are of course especially
interested in the relative motion of C1 and C2 and so our model
includes a fiducial relative position (at epoch 2000.0) as well as
relative proper motion in RA and DEC. Thus our model has six fit
parameters (if we include the core shift position angle as a fit
parameter, we do indeed obtain $46 \pm 1^\circ$, but prefer to fix
this via the larger scale jet axis). Our data set are the 13 $r$,
$\theta$ ($x$, $y$) position pairs over four epochs and four
frequencies, so the fit has $26-6=20$ degrees of freedom (DoF).

Using our measurements and the position error estimates above, we
performed a $\chi^2$ minimization to determine the model
parameters. These are listed in Table 4. The parameter error estimates
are somewhat subtle. Since the fit minimum has $\chi^2$/DoF = 2.78, we
must have systematic errors beyond those estimated above. The
conventional approach is to uniformly inflate all errors until the
effective $\chi^2$/DoF=1. This is equivalent to estimating errors
using the $\chi^2$ surface with increases of $+1,\, +2\, ... \times
\chi^2$/DoF. We list these ``1 $\sigma$'' and ``2 $\sigma$'' confidence
intervals in Table 4.

	However, this uniform inflation assumes that all errors have a Gaussian distribution and 
are equally underestimated. This is unlikely to be true. An alternative approach is to estimate
errors via a bootstrap analysis (Efron 1987). This has the virtue of using only the actual data values (not the error estimates),
but does pre-suppose that the observed data values are an unbiased draw
from an (unknown) error distribution about the true values.
Although our set of 13 position pairs is somewhat small for a robust bootstrap,
we generated 10,000 re-sampled realizations of the data set, replacing five pairs
in each realization with random draws from the remaining pairs. Each realization
was subject to the full least-squares fit for all model parameters. The histograms
of the fit values for the individual parameters were used to extract 95\% confidence intervals
for each quatity. These confidence intervals are listed in Table 4. They accord
fairly well with the inflated $\chi^2$ estimates.

	In general, the parameters appear well-constrained. The core-shift coefficients $a$ and $k$
are estimated to $\sim$3 \% accuracy (Table 4). The epoch position range is somewhat larger in the
bootstrap error analysis, evidently as a result of the substantial offset of the 2009.9
position from the general trend. 

The coefficient $k$ depends on the shape of electron energy spectrum, magnetic strength and particle density distribution \citep{lob98}. If $k$=1, it implies that the jet has a conical shape, where synchrotron self-absorption is the dominant absorption mechanism. We have obtained $k = 1.591\pm .232$ (via the bootstrap technique), in accordance with the literature (\citet{soko11}) where the highest reported $k$ value is $\thicksim$~1.5. This implies that our observations are consistent with the synchrotron self-absorption mechanism. 

The core-shift depends mainly on the frequency as well as the magnetic field and spectral index \citep{lob98}. All our measurements are derived from the same frequencies over time, however, there is a possibility of variation with magnetic field and spectral index. We calculated the spectral index for three epochs (2005, 2009 and 2015) using the peak intensities, and we have found these values to range from $-0.58$ to $-0.98$ and $-0.43$ to $-0.50$ for C1 and C2 respectively. From these ranges of spectral index, we can see that the variation over 10 years is quite small ($\thicksim 0.4$). To our knowledge, no time varying core-shift offsets have been reported in the literature. For our analysis, we assume that the core-shift is constant over time.

In Fig.~\ref{motions} we plot the relative C2 core position, shifted to infinite
frequency according to the best-fit $a$ and $k$, for each of the 13
observation frequencies and epochs. The plotted error
ellipses for the 8, 15 and 22 GHz observations are the formal $\sigma_x$ and $\sigma_y$ from the polarization analysis.  There are large outliers, especially the 2009.9 epoch. However, the overall shift is quite significant with motion detected in both RA
and DEC, at $>3\ \sigma$ in the $\chi^2$ analysis and at well over
$95$\% confidence in the bootstrap analysis. Additional epochs,
especially at high frequency will, however, be needed to make the motion
visually clear. 

To compare these above results, we also studied the motion of jet components. We find that the bright southern jet components continue to move away from the core, consistent with the previous results \citep{Rodri06}. For the weaker northern hotspot, the agreement is not as good, as it seems to exhibit inconsistent motion for some frequencies. However, this may be the result of errors in the previous measurements based on 5 GHz observations.

\begin{figure}
     \begin{center}
            \includegraphics[width=\textwidth, height = \textwidth]{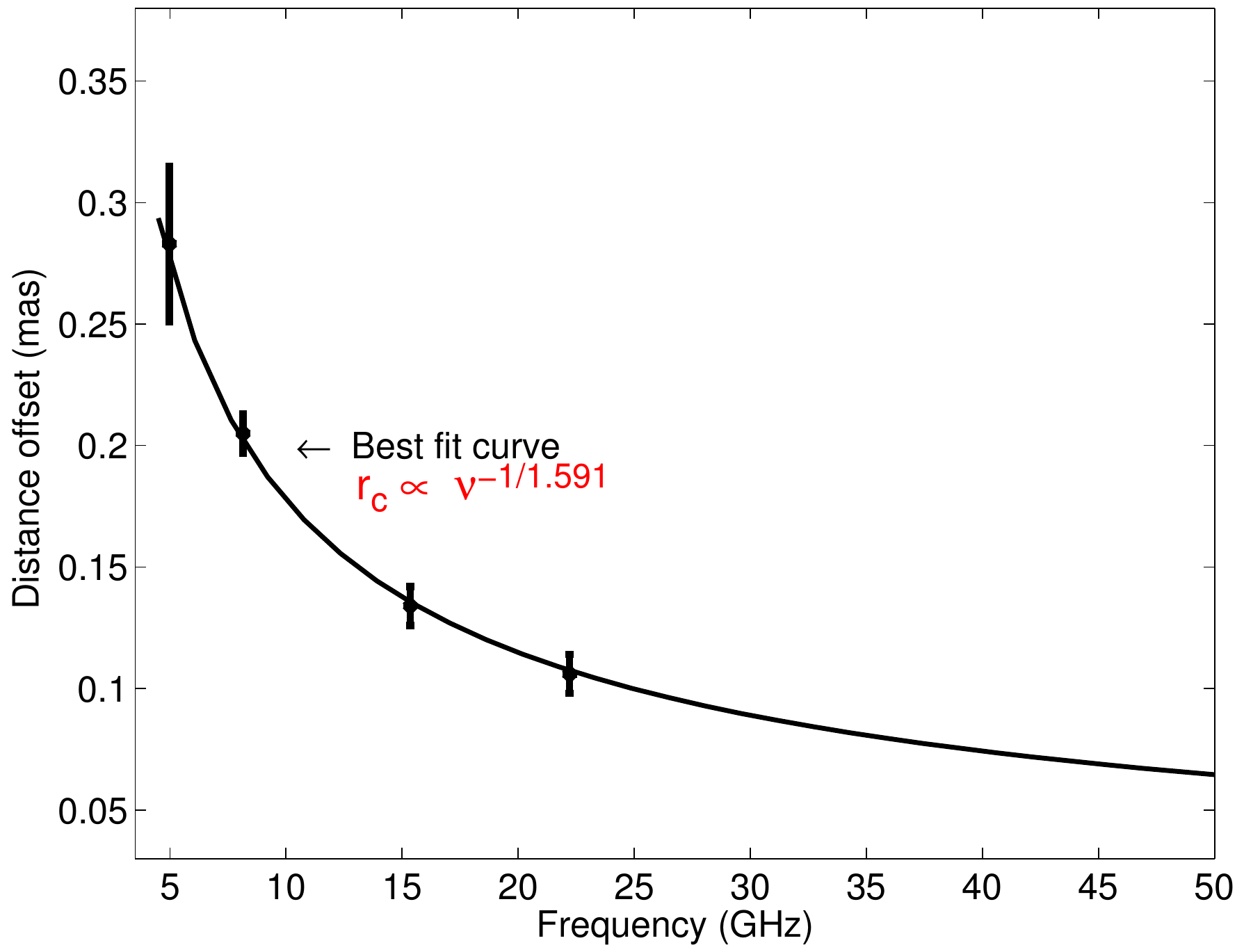}
    \end{center}
      \vspace*{0.0cm}
    \caption{Plot of the core-shift measurement in distance from the central engine for 0402+379 as a function of frequency. Black circles are observed distance offset from estimated infinite frequency core position at each frequency, and the black solid curve is the fitted function, with $r_c = a(\nu^{(-1/k)})$ (See Table 4).}
   \label{core-shift1}
\end{figure}

Using our best-fit $\mu_{RA}$ and $\mu_{Dec}$, we shift the raw data points (Table 3) to epoch 2000.0.  For each frequency, we obtain an average relative {\rm RA} and {\rm DEC}, which are subsequently subtracted from the fiducial 2000.0 point (Table 4) to obtain the distance from the core ($r_{c}$). This has been plotted to demonstrate our fitted frequency dependence, shown in Fig.~\ref{core-shift1}. The plotted errors have been obtained by error propagation using the above stated errors ($\sigma_x$ and $\sigma_y$).

\subsection{Magnetic Field Estimate}

The core-shift effect is useful to deduce various jet related physical parameters, including 
the magnetic field strength. \citet{lob98} and \citet{hiro}, for example provide a derivation that assumes
equipartition between the particle and magnetic field energy densities in the jet. An alternate formulation by \citet{Zd15} avoids the
equipartition assumption, using the flux density $F_{v}$ at a jet axis distance $h$
to estimate the magnetic field strength as
\begin{equation}
B_{F} (h) = \frac{3.35\times 10^{-11} \ D_{L} [{\rm pc}] \ \delta \ \Delta  \theta [{\rm mas}]^{5} 
\ {\rm tan}\Theta^{2}}{h[{\rm pc}] \ (\nu_{1}^{-1} - \nu_{2}^{-1})^{5} \ [(1+z) {\rm sin} \ i]^{3} \ F_{v}[{\rm Jy}]^{2}}
\end{equation}
where $z$ is redshift, $D_{L}$ is the luminosity distance in pc, $\delta = [\Gamma_{j}(1 \ - \ \beta_{j} cos \ i)]^{-1}$ 
is the Doppler factor, $\Gamma_{j}$ is the minimum Lorentz factor, $\beta_{j}$ is the jet bulk velocity factor (obtained from \cite{Rodri09}), $i$ is the inclination angle, $\Delta\theta$ is the observed angular core shift (Lobanov 1998), and $\Theta =$  arctan $\frac{\sqrt {d^{2} - b_{\phi}^{2}}}{2r}$ is the jet half opening angle \citep{pushkarev}. From the latest 8 GHz map, we have obtained the full width at half maximum of a Gaussian fitted to the transverse jet brightness component, $d = 4.130 \pm 0.017$ mas (minor axis of jet); the beam size along the jet direction, $b_{\phi} = 1.26$ mas; and the distance to the core along the jet axis, $r = 26.320\pm 0.017$ mas. Since the extended jet components are not readily detected at 15 and 22 GHz, we assume the same opening angle for all frequencies. 
Table~5 gives our estimated values for these parameters, with the
origins in the footnotes.  The numerical constant in the above equation has been obtained 
for $p=2$, where $p$ is index of the electron power law (see \citet{Zd12,Zd15}). 


From a weighted linear fit of  $\delta \theta$ against $\nu_{1}^{-1} - \nu_{2}^{-1}$, we obtain a slope  = $1.128 \pm 0.152$ and intercept  = $0.008 \pm 0.010$. Instead of calculating $\frac{\delta \theta}{\nu_{1}^{-1} - \nu_{2}^{-1}}$ for each frequency separately, its slope has been used in calculating the magnetic field strength. Our fit indicates a magnetic field strength $0.71 \pm 0.25 $ G at $h = 1$ pc, similar to that for other jets \citep{Sullivan}.


\subsection{Orbital Models}

	Our measured proper motion $\mu_{\rm RA}=-0.89 \pm 0.07\,\mu$as/y, $\mu_{\rm Dec}=1.29 \pm 0.10\, \mu$as/y
(symmetric width of the 95 \% CL bootstrap range) corresponds to a proper motion
$\mu$ of $1.57 \pm 0.08\, \mu$as/y at $PA_\mu=-34.6\pm 2.9^\circ$ (if we use the ``1 $\sigma$'' $\chi^2$ errors, the 
amplitude uncertainty is $\pm 0.38 \, \mu$as/y). Thus this is at least a $4\ \sigma$ detection.
It is consistent with the non-detection of a proper motion in Rodriguez et al. (2006), where 
15 years of 5 GHz data (1990-2005) were used to estimate $\mu = 6.7\pm 9.4 \,\mu$as/y; our higher frequency
data and core-shift correction are essential for measuring the much smaller motion.

We now ask if this proper motion is consistent with a shift due to the
relative orbits of the two BH. At $z=0.055$, it corresponds to a
projected space velocity of $\beta = v/c = 0.0054 \pm 0.0003$, so a
Keplerian analysis suffices.  First, the ratio $2\pi r/\mu = 2\pi \times 7.02\ {\rm mas}/0.00157$ mas/y = 28,000 y gives a characteristic orbital
timescale. Thus over our 12 y baseline, the core position PA has rotated
by less than a degree. This does not allow us to fit for precise
orbital parameters.  In particular, we have four measurements from the
VLBI analysis (relative position and proper motion) while we need six
parameters to define the relative orbit.

We note that the above derived $\sim 28000$ y period is rather close to the Earth's spin axis precession period of $\sim26000$ y, we believe this to be a coincidence. The differential astrometry performed here should not be expected by precession as that term has been removed with the correlator model and affects both the sources in an identical way.

\begin{figure}
     \begin{center}
            \includegraphics[width=1.0\textwidth, height = 1.0\textwidth]{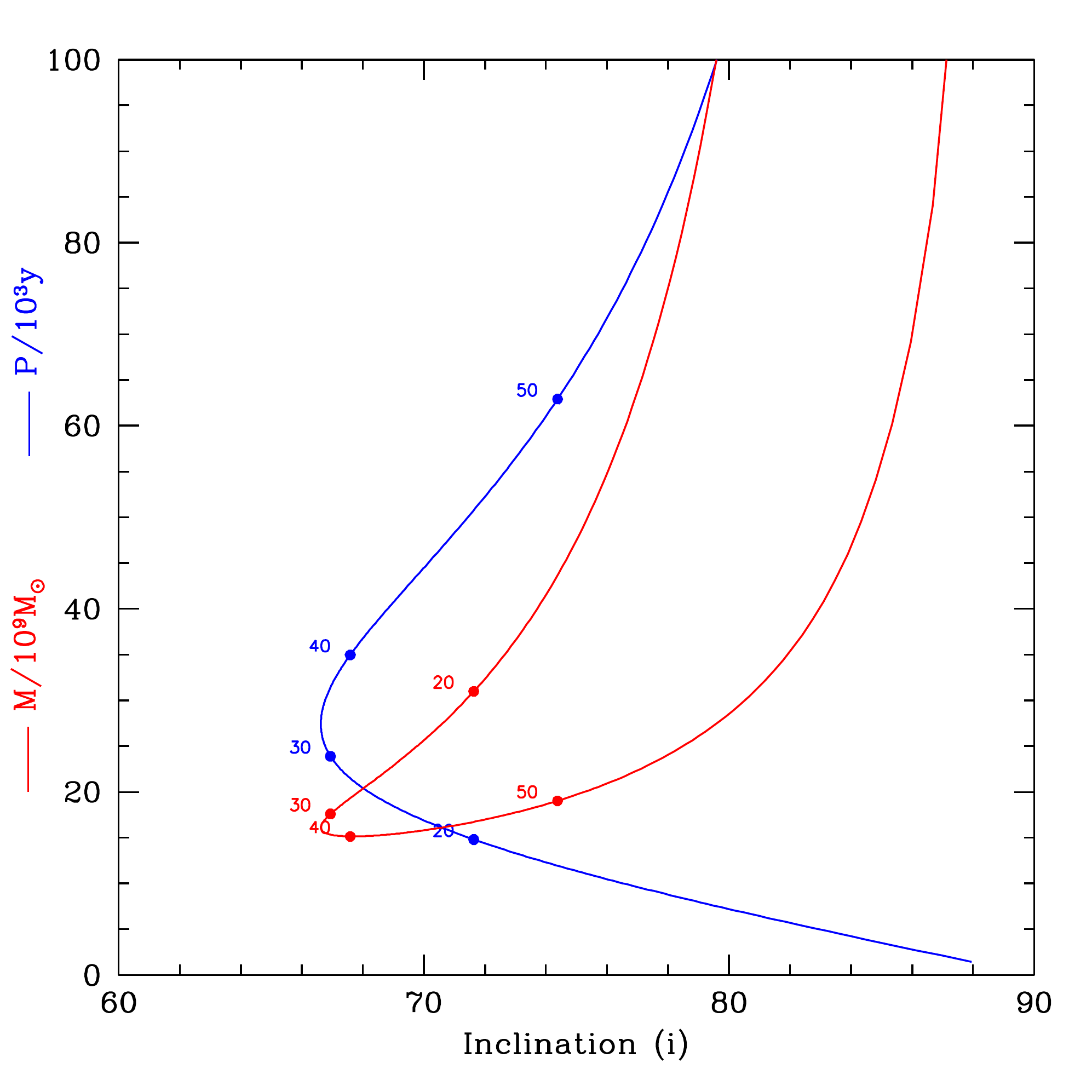}
    \end{center}
     \vspace*{-1.0cm}
    \caption{Orbital solutions for mass (red) and period (blue) as
a function of inclination angle. Points mark solutions with the projected PA given by the label numbers (in degrees North through East). }
   \label{orbsols}
\end{figure}

If we assume circular motion ($e=0$), then we can determine the
relative orbit in terms of one additional free parameter. In practice
it is easiest to select the PA of the projected orbit normal (measured
N through E) and then resolve the relative positions $x$ and $y$ (in
mas) and relative velocities $v_x$ and $v_y$ (in mas/y) in this
rotated coordinate system.  Then the orbit parameters are
$$
v= (v_x^2 - v_x v_y x/y)^{1/2}
$$
$$
a =  -(x^2 -x\, y\, v_x/v_y )^{1/2}
$$
$$
{\rm cos}(i) = [-y\,v_y /(x\,v_x)]^{1/2}
$$
$$
\theta = \pi+{\rm atan} \left ( [-y\,v_x/(x\,v_y)]^{1/2} - [1 -y\,v_x/(x\,v_y)]^{1/2}  \right )
$$
where $a$ and $v$ are the relative orbit radius and velocity, $i$ is the inclination and
$\theta$ gives the phase at our observation epoch. In fact it is more interesting to
plot the total mass $M=v^2 a/G$ and period $P=2\pi a/v$ against the orbit inclination $i$
(Fig.~\ref{orbsols}). Note that with our assumption of a circular orbit only fairly large inclinations
are consistent with our C1-C2 offset and relative motion (Fig.~\ref{motions}). Typical orbital periods are 
indeed 20-30ky, but the masses required by our apparent velocity are quite large 
$\ge 15 \times 10^9\ M_\odot$. With our nominal fit errors, the minimum mass is $M_9=15.4\pm 1.3$.
If one relaxes the $e=0$ assumption, smaller masses are allowed, but then the solution is 
nearly unconstrained.

The orbital eccentricity grows as the orbit shrinks since both stars and gas extract energy and angular momentum from the binary. For a stellar background, it depends on the mass ratio, with equal mass binaries producing orbits that are usually circular or slightly eccentric with $e< 0.2$ \citep{merritt07}. If a pair during the binary formation starts out with a non-zero eccentricity it may never become circular instead it tends to become more eccentric \citep{matsu11}. In the case of gas driven mergers, it depends on the disc thickness and the SMBBH’s location inside the disc. The critical value of e is reported to be $\sim 0.6$, such that system with high eccentricities tend to shrink to this value \citep{armi05, cuadra09,roedig12}. In the case of 0402+379, it has been found to be embedded in cluster gas \citep{andra16}, which makes it likely to have a non-zero eccentricity.

\begin{figure}
     \begin{center}
            \includegraphics[width=1.0\textwidth, height = 1.0\textwidth]{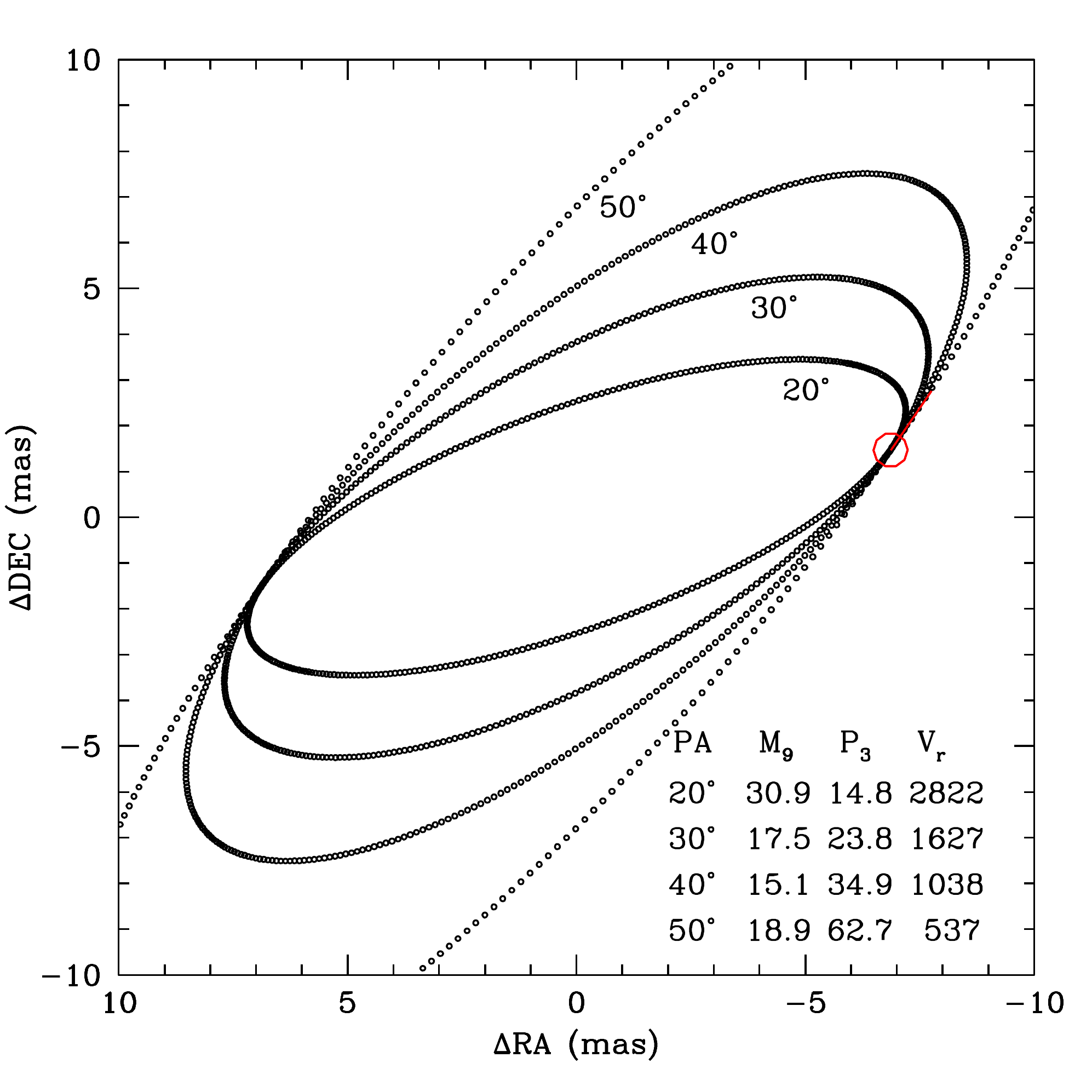}
    \end{center}
     \vspace*{-1.0cm}
    \caption{Circular orbit fits for the four PA values marked in figure 5. All are consistent with the
observed offset and proper motion (red). The mass, period and relative radial velocity for
the solutions for each PA value are listed in the figure.}
   \label{4orb}
\end{figure}

In \citet{Rodri09} HI absorption measurements were used to infer kinematic motion 
about an axis inclined $\sim 75^\circ$ to the Earth line of sight, passing through C2,
the origin of the kpc-scale jets. If we look at the solution derived here we see
that the $PA =47^\circ$ axis corresponds to $i=71.3^\circ$ (Fig.~\ref{orbsols} \& Fig.~\ref{4orb} ), in reasonable agreement with the HI estimate. The binary spin can be different from the orbital angular momentum, however, it’s been found that if the amount of gas accreted is high (1-10\% of the black hole) on the timescales of binary evolution, it can change according to the orbital axis (\citet{schnit13}, and references therein). Binary orbital axis and individual black hole spins tend to realign due to interaction with external gas except when the mass ratios are extreme ($>>1$) whereas, torques from stars can cause misalignments of the binary orbit orientation from the disc \citep{miller13}. For this fit we have $P=49$ ky and $M=16.5 \times 10^9\ M_\odot$.

	Because we find a large proper motion $\mu$ we expect the orbital motion
to induce a substantial radial velocity in the relative orbit. Some values are given 
in Fig.~\ref{4orb} and for $PA=47^\circ$ we expect a relative $v_r=700$ km/s. While the HI
measurements do show velocity differences of this order, we do not see such large
velocities in the optical line peaks. Examining the Keck spectra in \citet{romani14}
we see that the stellar features of the elliptical host center on 
$16,618\pm 53$ km/s, while the Seyfert I-type narrow-line core emission 
centers on $16,490$ km/s.
Narrow line emission extends several arcsec from the core spanning $\sim 300$ km/s
while in the unresolved kpc core the velocity dispersion is $\sim 750$ km/s.
Thus, while at least $2\times 10^{10}\ M_\odot$ lies within the central kpc, we do not
see multiple components shifted by $>500$ km/s. However, the full line width does
accommodate such velocities and the wings of the H$\alpha$ complex are centered
at $\sim 17,020$ km/s suggesting that fainter broad line emission might include
components spread over $>1000$ km/s. Further $v_r$ above is the relative velocity;
if only the heavier component has bright optical emission, then the broad line velocity
shift from the background galactic velocity (arguably near the center of mass velocity)
will be reduced to $m v_r/M_{Tot}$. Indeed, \citet{Rodri09} assume that the jet-producing
C2 core is the dominant mass and the center of rotation. If this core also dominates
the broad line emission, then we expect that our VLBI relative velocity is dominated by
the motion of C1 and the optical radial velocity shift from the host velocity may be small.

	We must also compare with other core mass estimates. As noted above the optical
lines indicate several $\times 10^{10}\ M_\odot$ in the central kpc.  HI absorption
velocities require $>7\times10^{8}\ M_{\odot}$ \citep{Rodri09}. And finally the
host bulge luminosity also indicates a large hole mass $M_\bullet \sim 3 \times 10^9\ M_\odot$
\citep{romani14}. All of these suggest substantial hole mass. The very large
$M_9 \sim 15$ masses implied by our fits are not excluded but do stretch the available
mass budget. 

\subsection{Comments on the resolved SMBHB Population}

	The process of hierarchical merging should make close SMBHB common, but to date few 
candidates at sub-kpc separations have been seen. The resolved (massive) SMBHB seem to 
be preferentially in galaxy clusters or their products.
For example the SMBHB candidate RBS 797 \citep{git13} resides in a cool-core cluster at z=0.35. 
0402+379 itself lies in a massive galaxy and dense X-ray halo (likely a fossil cluster) at z = 0.055. So 
such environments seem to be a good place to look for additional systems. Another path to discovering
multi-BH nuclei has been described by \citet{dean14} who find J1502+1115 to be a {\it triple} system,
with a closest pair separation of 140 pc at redshift z=0.39. Compact radio jets in the closest 
pair of this source exhibit rotationally symmetric helical structure, plausibly due to binary-induced 
jet precession. However the total number of resolved compact cores at pc scales seems very small
with 0402+379 remaining the only clear example out of several thousand mapped sources \citep{burke,tremb}.

	Of course systems of even smaller separation are of the greatest interest since at
$r \sim 0.01$ pc losses from gravitational radiation will dominate and the merging binaries
can be an important signal in pulsar timing studies \citep{ravi}. At present, we
rely on arguments about evolution of the wider systems to infer the existence of merging SMBHB.
If such evolution occurs we might hope for a discovery of an intermediate $r \sim 0.1$ pc scale 
massive $>10^9\ M_\odot$ system at low $z$ where sufficient resolution for a kinematic 
binary study is possible with high frequency VLBI. Such a binary would have $P<10^3$ y and a 
well-constrained visual orbit should be achievable, making possible a precision test of the SMBHB nature \citep{trio14}.
However, we should note that our study of the galactic halo of 0402+379 \citep{andra16} implies
that it has stalled at its present 7 pc separation for several Gyr, so the path between
resolvable and gravitational radiation-dominated SMBHB may not always be smooth.

\section{Conclusion}

In this study of 0402+379, we have focused on two aspects: frequency dependent core-shift 
and secular relative core motion. Both effects are observed, but the measured values present 
interpretation challenges.

The strong observed core-shift matches well with the large-scale jet axis. It
also provides quite typical estimates for the jet base magnetic field of $\sim 0.45-0.95$\, G.  
However the core-shift index 1.591 (1.556-1.823 95\% CL range) is somewhat large (expected $k \sim 1$), with the
highest reported value in literature is $k\thicksim$~1.5 \citep{soko11}.
This may be an artifact of our constant core-shift fit assumption, as perturbations could
arise from the epoch-to-epoch variation in the underlying core component fluxes.

	After accounting for this core shift the infinite frequency relative positions 
of the C1 and C2 cores undergo a statistically significant secular proper motion.
The motion corresponds to $\beta = 0.0054 \pm 0.0003$ and, if orbital, it represents
the first direct detection of orbital motion in a SMBHB, and promotes this system to a visual binary.
Although we do not have sufficient observables to solve for an orbit, we can find
plausible orbits, even assuming $e=0$. Intriguingly such orbits align well with the
large scale jet axis and have similar inclination to those estimated with HI absorption
VLBI. But the required masses are quite large (highest reported mass is 21 billion $M_{\odot}$, \cite{mcconnell}).
To test our orbital picture, additional VLBI epochs to
confirm the consistency of the proper motion will be essential, and further studies of
the core dynamics, especially at sub-kpc scales will also be very important. We should not
forget that including a finite orbital eccentricity can allow smaller masses, but
we will need additional kinematic constraints to motivate such solutions.

	Thus discovery of possible orbital motion in 0402+379 presents the exciting prospect
of probing a SMBHB's kinematics. Certainly,
extensions to our high frequency VLBI campaign can improve the measurements, but this 
proper motion is perhaps the most exciting as a spur to searches for tighter, faster and more easily measured
examples of resolved SMBHBs.

{\bf Acknowledgements:}

This research has made use of NASA's Astrophysics Data System. English
translations of Einstein (1916) and Einstein (1918) made available via
The Digital Einstein Papers of the Princeton University Press aided
the preparation of this paper. The National Radio Astronomy
Observatory is a facility of the National Science Foundation operated
under cooperative agreement by Associated Universities, Inc..  Partial
support for this work was provided by the National Aeronautics and
Space Administration through Chandra Award Number GO4-15121X issued by
the Chandra X-ray Observatory Center, which is operated by the
Smithsonian Astrophysical Observatory for and on behalf of the
National Aeronautics Space Administration under contract NAS8-03060.

\clearpage

\begin{table}
\begin{center}
Table 1: Observations \\
\vspace{0.2cm}
\begin{tabular}{lcccccc}
\hline
\hline
Frequency & Date & Integration & BW & Polarization & IF & Reference\\
(GHz)& & time (min)& (MHz) & & &\\
\hline
4.98 		 & 01/24/2005 & 69   &   8 & 2 & 4 & \cite{Rodri06}\\
4.98		 & 12/28/2009 & 286 &  32 & 4 & 4 & This paper\\
4.98		 & 06/20/2015 & 70   & 32 & 4 & 8 & This paper\\
8.15            & 06/13/2005 & 69   & 8   & 2  & 4  & \cite{Rodri06} \\
8.15            &12/28/2009 & 261& 32 & 4 & 8 & This paper \\
8.15            &06/20/2015 & 261& 32 & 4 & 8  & This paper \\ 
15.35          &03/02/2003 & 478& 16 & 2 & 4&  \cite{man03}\\
15.35          &01/24/2005 & 122&  8  & 2 & 4  & \cite{Rodri06}\\
15.35          &12/28/2009 & 292& 32 & 4 & 8& This paper\\
15.35          &06/20/2015 & 286& 32 & 4 & 8  & This paper\\ 
22.22          &06/13/2005 & 251&  8  & 2 & 4  & \cite{Rodri06}\\
22.22          &12/28/2009 & 325& 32 & 4  & 8 & This paper    \\
22.22          &06/20/2015 & 334& 32 & 4  & 8 & This paper\\ 
\hline
\end{tabular}
\label{Obs}
\end{center} 
\end{table}
 
\clearpage

\begin{table}
\begin{center}
Table 2: Stationary Gaussian Model Components \\
\vspace{0.2cm}
\begin{tabular}{lcccccc}
\hline
Frequency & $a$(C1) & $b/a$(C1)& $\phi$(C1) & $a$(C2) &$b/a$(C2)& $\phi$(C2) \\
(GHz)& (mas) & & ($^o$)& (mas) & & ($^o$)\\
\hline
5 & 0.563 & 0.000 & 82.80 &1.270 & 0.130 & 6.60 \\
8 & 0.451 & 0.420 & 74.00 & 0.420 & 0.490 & 8.60\\
15 & 0.249 & 0.360 & 77.00 & 0.230 & 0.000 & 21.40\\
22 & 0.218 & 0.160 & 78.90 & 0.170 & 0.390 & 27.80 \\
\hline
\end{tabular}
\label{fixed}
\end{center}
{Fixed model parameters of Gaussian components for C1 and C2 of the model brightness distribution at each frequency. These are: $a$, semi-major axis; $b/a$, axial ratio (where $b$ is semi-minor axis); $\Phi$, component orientation for both C1 and C2. All angles are measured from North through East. \\}
\end{table}

\clearpage

\begin{table}
\begin{center}
Table 3: Variable Gaussian Model Components \\
\vspace{0.2cm}
\begin{tabular}{lccccc}
\hline\hline
Epoch & Frequency & $S_{\nu}(C1)$ & $S_{\nu}(C2)$ & $r$ & $\theta$ \\
& (GHz) &  (Jy) & (Jy) & (mas) & ($^o$)\\
\hline
2005.07 & 5 & 0.057 $\pm$ 0.005 & 0.014 $\pm$ 0.001 & 6.942 & -75.70  \\
2009.99 & 5 & 0.058 $\pm$ 0.005 & 0.016 $\pm$ 0.001 & 6.841 & -75.93  \\    
2015.43 & 5 & 0.060 $\pm$ 0.006 & 0.013 $\pm$ 0.001& 6.884 & -75.79  \\
2005.45 & 8 & 0.067 $\pm$ 0.006 & 0.016 $\pm$ 0.002 & 6.913 & -76.46  \\
2009.99 & 8 & 0.052 $\pm$ 0.004 & 0.017 $\pm$ 0.002 & 6.920 & -76.42  \\
2015.43 & 8 & 0.083 $\pm$ 0.008 & 0.018 $\pm$ 0.002 & 6.913 & -76.33  \\
2003.17 & 15 & 0.070 $\pm$ 0.007 &  0.020 $\pm$ 0.002 & 6.929 & -76.96  \\
2005.07 & 15 & 0.054 $\pm$ 0.005 & 0.016 $\pm$ 0.002 & 6.959 & -76.81  \\
2009.99 & 15 & 0.029 $\pm$ 0.003 & 0.012 $\pm$ 0.001 & 6.985 & -76.96 \\
2015.43 & 15 & 0.058 $\pm$ 0.005 & 0.015 $\pm$ 0.001 & 6.956 & -76.77  \\
2005.45 & 22 & 0.037 $\pm$ 0.003 & 0.011 $\pm$ 0.001& 6.950 & -77.08  \\
2009.99 & 22 & 0.020 $\pm$ 0.002 & 0.011 $\pm$ 0.001& 6.984 & -77.16   \\
2015.43 & 22 & 0.040 $\pm$ 0.003 & 0.012 $\pm$ 0.001 & 6.969 & -77.04  \\
 \hline\\
\end{tabular}
\label{variable}
\end{center}
{Variable model parameters of Gaussian components for C1 and C2 of the model brightness distribution at different epoch and frequency. These are as follows: $S_{\nu}$, flux density at each frequency; $r$, $\theta$, polar coordinates of the center of the component C2 relative to the center of component C1 (it has been assumed to be at a fixed position). Errors in flux have been estimated using both flux systematics and map rms ($\sqrt((0.1*S_{\nu})^2+rms^2)$).} 
\end{table}

\clearpage


\begin{table}
\begin{center}
Table 4: Fitting Parameters \\
\vspace{0.2cm}
\begin{tabular}{lccccc}
\hline
\hline
\multirow{2}{*}{} & \multirow{2}{*}{} & {Technique} & \multicolumn{2}{c}{$\chi^{2}$}\\

 Parameters & Value & Bootstrap (95\% )& $1\ \sigma$ &  $2\ \sigma$\\
\hline
$\Delta{\rm RA}_0$(mas) &  -6.863 & -6.892, -6.855 &  -6.859, -6.868  &  -6.858, -6.869\\
$\Delta{\rm DEC}_0$(mas)  & 1.474 & 1.448, 1.478 &  1.470, 1.478  & 1.468, 1.480\\
$\mu_{RA}$($\mu$as/y) & -0.887 & -0.970, -0.831& -1.245, -0.549 & -1.389, -0.405\\
$\mu_{DEC}$($\mu$as/y) & 1.286 & 1.200, 1.401 & 0.878, 1.672 & 0.713, 1.836\\
a (mas) & 0.756 & 0.700, 0.818 & 0.739, 0.777 & 0.731,  0.785\\
k & 1.591 & 1.556, 1.823 & 1.565, 1.617 & 1.555, 1.628\\
\hline\\
\end{tabular}
\label{variable}
\end{center}
{Fitted parameters values (Column 2) and their corresponding confidence intervals obtained from two different technique: bootstrap analysis (Column 3) and $\chi^{2}$ minimization (Column 5 $\&$ 6). $\Delta{\rm RA}_0$ and $\Delta{\rm DEC}_0$ are infinite frequency core offsets at epoch 2000.0; $\mu_{RA}$ and $\mu_{DEC}$ are proper motion estimates; $a$ and $k$ are core-shift fitting parameters ($r_{c} = a\ \nu^{(-1/k)}$).} 
\end{table}

\clearpage

\begin{table}
\begin{center}
Table 5:  C2 Jet parameters\\
\vspace{0.2cm}
\begin{tabular}{lcccccc}
\hline\hline
Redshift  & Luminosity & Half opening & Bulk Velocity & Inclination & Lorentz & Doppler Factor \\
$z$ & Distance $D_{L}$ (Mpc) & angle $\Theta$ ($^\circ$) & factor $\beta_{app}$ & angle $i$ ($^\circ$) & factor $\Gamma_{j}$ & $\delta$\\
\hline
0.055 & 242.2 \tablenotemark{a} & 4.29 \tablenotemark{b} & 0.4 \tablenotemark{c} & 71.3 \tablenotemark{d} &1.077 \tablenotemark{e} &1.11 \tablenotemark{f}\\
\hline
\end{tabular}

\label{jet}
\end{center}
{\tablenotemark{a}~Luminosity distance was obtained for cosmological model : $H_{0}=71$ kms$^{-1}$Mpc$^{-1}$,  $\Omega_{\Lambda}$ = 0.73,  $\Omega_{M}$ = 0.27}\\
{\tablenotemark{b}  \citet{pushkarev}}\\
{\tablenotemark{c} \citet{Rodri09}.}\\
{\tablenotemark{d} Section 3.1.}\\
{\tablenotemark{e} Lorentz factor, $\Gamma_{j} = (1 \ + \ \beta_{app}^2)^{\frac{1}{2}}$ \citep{Zd15}.}\\
{\tablenotemark{f} Doppler factor, $\delta = [\Gamma_{j}(1 \ - \ \beta_{j} {\rm cos} \ i)]^{-1}$ \citep{Zd15}.}\\
\end{table}					

\clearpage

\clearpage

\clearpage


\clearpage

\clearpage





\begin{thebibliography}

\bibitem[Abbott et al.(2016)]{gw16} Abbott, B.~P.,  Abbott, R.,  Abbott, T.~D., Abernathy, M.~R.  et al.  2016, Physical Review Letters, 116, 6, 061102

\bibitem[Amaro-Seoane et al.(2013)]{elisa} Amaro-Seoane, P.,  Aoudia, S., Babak, S., et al., 2013, GW Notes, 6, 4

\bibitem[Andrade-Santos et al.(2016)]{andra16} Andrade-Santos, F., Bogdan, A., Romani, R.W., et al.  2016, ApJ, 826, 91

\bibitem[Armitage \& Natarajan(2005)]{armi05}Armitage, P.~J. and Natarajan, P., \apj, 2005, 634, 921

\bibitem[Arzoumanian et al.(2016)]{arzou16} Arzoumanian, Z., Brazier, A., Burke-Spolaor, S., et al., 2016, \apj, 821, 13 

\bibitem[Barnes(2002)]{barnes}Barnes, J. E., \mnras, 2002, 333, 481

\bibitem[Begelman et al.(1980)]{beg80} Begelman, M.~C., Blandford, R.~D., \& Rees, M.~J. 1980, Nature, 287, 307

\bibitem[Berentzen et al.(2009)]{berent09}Berentzen, I. and Preto, M. and Berczik, P. et al., 2009, \apj, 695, 455
\bibitem[Blandford et al.(1979)]{bland79}Blandford, R. D. \& Konigl, a. 1979, The Astrophysical Journal, 232,

\bibitem[Burke-Spolaor(2011)]{burke} Burke-Spolaor, S.  2011, MNRAS, 410, 2113

\bibitem[Callegari et al.(2009)]{cal09}Callegari, S. and Mayer, L. and Kazantzidis, S. et al., \apjl, 2009, 696, L89

\bibitem[Callegari et al.(2011)]{cal11}Callegari, S. and Kazantzidis, S. and Mayer, L., et al., \apj, 2011, 729, 85

\bibitem[Cuadra et al.(2009)]{cuadra09}Cuadra, J. and Armitage, P.~J. and Alexander, R.~D. and Begelman, M.~C., \mnras, 2009, 393,1423
\bibitem[Deane et al.(2014)]{dean14} Deane, R. P., Paragi, Z.,  Jarvis, M. J., et al.  2014 
Nature, 511, 57

\bibitem[Dotti et al.(2007)]{dotti07}Dotti, M. and Colpi, M. and Haardt, F. et al., 2007, \mnras, 379, 956

\bibitem[Dotti et al.(2012)]{dotti12}Dotti, M. and {Sesana}, A. and {Decarli}, R., Advances in Astronomy, 2012, 2012, 940568
\bibitem[Einstein(1916)]{ae16} Einstein, Albert 1916, 
Sitzungsberichte der K{\"o}niglich Preu{\ss}ischen Akademie 
der Wissenschaften, 688

\bibitem[Einstein(1997)]{ae97} Einstein, Albert 1997,  
The Collected Papers of Albert Einstein, Volume 6:
The Berlin Years: Writings, 1914-1917 (English translation supplement) 
Translator: Alfred Engel; Consultant: Engelbert Schucking 
(Princeton, NJ: Princeton University Press) 201

\bibitem[Einstein(1918)]{ae18} Einstein, Albert 1918, 
Sitzungsberichte der K{\"o}niglich Preu{\ss}ischen Akademie der 
Wissenschaften, 154

\bibitem[Einstein(2002)]{ae02} Einstein, Albert 2002, 
The Collected Papers of Albert Einstein, Volume 7:
The Berlin Years: Writings, 1918-1921 (English translation supplement) 
Translator: Alfred Engel; Consultant: Engelbert Schucking 
(Princeton, NJ: Princeton University Press) 9

\bibitem[Escala et al.(2004)]{escala04}Escala, A. and Larson, R.~B. and Coppi, P.~S. et al., 2004, {\apj}, 607, 765

\bibitem[Escala et al.(2005)]{escala05}Escala, A. and Larson, R.~B. and Coppi, P.~S. et al., 2005, {\apj}, 630, 152

\bibitem[Fomalont(1999)]{fomalont} Fomalont E.~B., Image Analysis, Synthesis Imaging in Radio Astronomy II, 1999, 301

\bibitem[Gaskell(2010)]{gask10} Gaskell, C.~M.  2010, \nat, 463, E1

\bibitem[Gitti et al.(2013)]{git13} Gitti, M \& Giroletti, M \& Giovannini, G \& 
Feretti, L \& Liuzzo, E.  2013, \aap, 557, L14

\bibitem[Hirotani(2005)]{hiro} Hirotani, K., 2005, \apj, 619, 73






\bibitem[Hartkopf et al.(2001)]{hmw01}Hartkopf, W.I. and Mason, B.~D. and Worley, C.E., \aj, 2001,  122, 3472

\bibitem[Hayasaki(2009)]{hayasaki09}Hayasaki, K., 2009, \pasj, 61, 65

\bibitem[Heintz(1978)]{h78} Heintz, W. D. 1978, Double Stars (Reidel, Dordrecht), 63

\bibitem[Khan et al.(2011)]{khan11}Khan, F.M. and Just, A. and Merritt, D., \apj, 2011, 732, 89
\bibitem[Khan et al.(2016)]{khan16}Fazeel Mahmood Khan, Davide Fiacconi, Lucio Mayer et al., 2016, \apj, 828, 2

\bibitem[Klein et al.(2016)]{elisa2}Klein, A., Barausse, E., Sesana, A., et al., 2016, \prd, 93, 024003

\bibitem[Lobanov(1998)]{lob98} Lobanov, A.~P.  1998, \aap, 330, 79


\bibitem[Maness et al.(2004)]{man03} Maness, H.~L., Taylor, G.~B., Zavala, R.~T., et al., 2004, \apj, 602, 123 

\bibitem[Matsubayashi et al.(2011)]{matsu11}Matsubayashi, T. and Makino, J. and Ebisuzaki, T., 2007, \apj, 656, 879

\bibitem[McGary et al. (2001)]{mcgary01}McGary, R.~S. and Brisken, W.~F. and Fruchter, A.~S. et al., \aj, 2001, 121, 1192
\bibitem[McConnell et al.(2011)]{mcconnell}McConnell, N.~J. and Ma, C.-P. and Gebhardt, K. et al., \nat, 2011, dec, 480, 215

\bibitem[Merritt et al.(2005)]{merritt05}Merritt, D. and Milosavljevi{\'c}, M., 2005, Living Reviews in Relativity, 8

\bibitem[Merritt(2006)]{merritt06}Merritt, D., 2006, Reports on Progress in Physics, 69, 2513

\bibitem[Merritt et al.(2007)]{merritt07}Merritt, D. and Mikkola, S. and Szell, A., 2007, \apj, 671, 53

\bibitem[Merritt et al.(2011)]{merritt11}Merritt, D. and Vasiliev, E., \apj, 2011, 726, 61

\bibitem[Milosavljevi{\'c} \& Merritt(2003)]{mm03} Milosavljevi{\'c}, M., \& Merritt, D.\ 2003, \apj, 596, 860

\bibitem[Coleman Miller and Krolik(2013)]{miller13} Coleman Miller, M. and Krolik, Julian H., 2013, \apj, 774, 43
\bibitem[Morganti et al.(2009)]{Morganti} Morganti, R.,  Emonts, B. \& Oosterloo, T.  2009, \astro, 496, L9

\bibitem[Pollack et al.(2003)]{pol03} Pollack, L.~K., Taylor, G.~B.\& Zavala, R.~T.  2003, \apj, 589, 733

\bibitem[Pushkarev et al.(2012)]{pushkarev} Pushkarev , A.~B. \& Lister, M.~L. \& Kovalev, Y.~Y. \& Savolainen, T., May, 2012, ArXiv e-prints,1205.0659

\bibitem[Richstone et al.(1998)]{rich98} Richstone, D., Ajhar, E. A., Bender, R. et al. 1998, \nat, 395, A14 

\bibitem[Ravi et al.(2015)]{ravi} Ravi, V., Wyithe, J.~S.~B.,  Shannon, R.~M. et al. \mnras, 2015, 447, 2772

\bibitem[Roberts et al.(1991)]{roberts91}Roberts, D.~H. and Brown, L.~F. and Wardle, J.~F.~C., IAU Colloq. 131: Radio Interferometry. Theory, Techniques, and Applications, 1991,19, 281

\bibitem[Rodriguez et al.(2006)]{Rodri06} Rodriguez, C., Taylor, 
G.~B., Zavala, R.~T. et al.  2006, \apj, 646, 49

\bibitem[Rodriguez et al.(2009)]{Rodri09} Rodriguez, C., Taylor, G.~B., Zavala, R.~T. et al. 2009, \apj, 697, 37

\bibitem[Roedig et al.(2012)]{roedig12} Roedig, C. and Sesana, A. and Dotti, M. et al., 2012, \aap, 545, A127

\bibitem[Romani et al.(2014)]{romani14} Romani, R.~W., Forman, W.~R., Jones, C. et al.  2014, ApJ, 780, 149

\bibitem[Schnittman(2013)]{schnit13} Schnittman, J.~D., Classical and Quantum Gravity, 2013, 30, 24

\bibitem[Sesana et al.(2006)]{sesana06}Sesana, A. and Haardt, F. and Madau, P., \apj, 2006, 651, 392

\bibitem[Sesana et al.(2007)]{sesana07}Sesana, A. and Haardt, F. and Madau, P., \apj, 2007, 660, 546

\bibitem[Shepherd et al.(1995)]{Shep95} Shepherd, M. C., 
Pearson, T.J., \& Taylor, G.B.  1995, BASS, 27, 903

\bibitem[Shannon et al.(2015)]{shan15}Shannon, R.~M.  Ravi, V., Lentati, L.~T. et al. Science, 2015, 349, 1522

\bibitem[Sokolovsky et al.(2011)]{soko11} Sokolovsky, K.~V., Kovalev, Y.~Y., Pushkarev, A.~B. et al.  2011, \aap, 532, A38

\bibitem[Stewart et al.(2009)]{stew09} Kyle R. Stewart, James S. Bullock, Risa H. Wechsler et al. 2009, \apj, 702, 1

\bibitem[O'Sullivan el al.(2009)]{Sullivan} O'Sullivan, S.~P. \& Gabuzda, D.~C., \mnras, 2009, 400, 26

\bibitem[Taylor et al.(1996)]{Taylor96} Taylor, G.~B., Readhead, A.~C.~S., \& Pearson, T.~J.  1996, \apj, 463, 95 

\bibitem[Taylor(2014)]{trio14} Taylor, G., \nat, 2014, 511, 35

\bibitem[Tremblay et al.(2016)]{tremb} Tremblay, S.E., Taylor, G.B., Ortiz, A.A., et al., 2016, MNRAS, p. stw592 (Submitted)

\bibitem[Ulvestad et al.(2001)]{Ulvestad01} Ulvestad, J.,
Greisen, E.~W. \& Mioduszewski, A.\ 2001, AIPS Memo 105:AIPS Procedures
for initial VLBA Data Reduction, NRAO

\bibitem[van Moorsel et al.(1996)]{van96} Van 
Moorsel, G., Kemball, A., \& Greisen, E.  1996, ASP Conf.~Ser.~101: 
Astronomical Data Analysis Software \& Systems V, 5, 37 

\bibitem[Xu et al.(1995)]{Xu95} Xu, W., Readhead, A.~C.~S., 
Pearson, T.~J., Polatidis, A.~G., \& Wilkinson, P.~N. 1995, \apjs, 99, 297 

\bibitem[Zdziarski et al.(2012)]{Zd12} Zdziarski, A.~A. \& Lubi{\'n}ski, P. \& Sikora, M., \mnras, 2012, 423, 663

\bibitem[Zdziarski et al.(2015)]{Zd15} Zdziarski, A.~A. \& Sikora, M. \& Pjanka, P. \& Tchekhovskoy, A.\mnras, 2015, 451, 927


\end{thebibliography}
\end{document}